# Linear Polarization Measurements for High-Spin States in $^{146}$Gd

Krishichayan[1*], Rajashri Bhattacherjee[2], S.K. Basu[3], R.K. Bhowmik[4], A. Chakraborty[5], L. Chaturvedi[6], A. Dhal[7], U. Garg[8], S.S. Ghugre[2], R. Goswami[2], A. Jhingan[4], N. Madhvan[4], P.V. Madhusudhana Rao[9], S. Mukhopadhyay[10], S. Muralithar[4], S. Nath[4], N.S. Pattabiraman[2], S. Ray[1], S. Saha[1], M. Saha Sarkar[1], S. Sarkar[11], R. Singh[12], R.P. Singh[4], A.K. Sinha[2], R.K. Sinha[7], P. Sugathan[4], B.K. Yogi[13]

*[1]Saha Institute of Nuclear Physics, I/AF, Bidhan Nagar, Kolkata 700064, India*

*[2]UGC-DAE Consortium for Scientific Research, III/LB-8, Bidhan Nagar, Kolkata 700098, India*

*[3]Variable Energy Cyclotron Centre, I/AF, Bidhan Nagar, Kolkata 700064, India*

*[4]Inter University Accelerator Centre, Aruna Asaf Ali Marg, New Delhi 110067, India*

*[5]Department of Physics, Krishnath College, Berhampore 742101, India*

*[6]Guru Ghasidas Vishwavidyalaya, Bilaspur 495009, India*

*[7]Department of Physics, Banaras Hindu University, Varanasi 221005, India*

*[8]Department of Physics, University of Notre Dame, Notre Dame, IN 46556, USA*

*[9]Jnanananda Laboratories for Nuclear Research, Department of Nuclear Physics, Andhra University, Visakhapatnam 530003, India*

*[10]Nuclear Physics Division, Bhabha Atomic Research Centre, Mumbai 400085, India*

*[11]Department of Physics, Bengal Engineering and Science University, Shibpur, Howrah 711103, India*

*[12]Department of Physics and Astrophysics, University of Delhi, New Delhi 110007, India*

*[13]Department of Physics, Govt. College, Kota 324009, India*

---

*correspondence author: *krishichayan@gmail.com*

**Abstract**

A γ-ray linear polarization measurement has been performed to directly determine the parities for the levels in $^{146}$Gd nucleus. High-spin states in this nucleus were populated in a reaction $^{115}$In + $^{34}$S at 140 MeV incident energy. Linearly polarized γ - rays emitted from oriented states were measured using a Compton polarimeter consisting of an array of 8 Compton-suppressed Clover detectors. Unambiguous assignments of the spin and parity have been made for most of the observed levels and changes made in the previously reported spin-parity assignments for a few levels. Shell model calculations performed with judicious truncation over the π(*gdsh*) valence space interpret the structure of only the low-lying levels up to $J^{\pi} = 19^{+}$ and $9^{-}$. $N = 82$ neutron-core breaking is found to be essential for high spin states with excitation energies $E_x > 7$ MeV.

## I. INTRODUCTION

The nuclei in mass $A \sim 150$ region have drawn considerable attention, both experimentally and theoretically due to their proximity to the $N = 82$ and $Z = 64$ shell closures. Generation of angular momentum in these nuclei has been a subject of interest for a long time. The low-lying region of the excitation spectra for these nuclei show irregular and complex patterns, typical of near-spherical nuclei and are dominated by single- and multi-particle excitations. The unprecedented advances in experimental techniques and facilities in recent years have made it possible to study these nuclei under extreme conditions of angular momentum. Such studies produce wealth of data revealing many new aspects of nuclear structure, hitherto unknown. The new experimental information on nuclei close to the shell closures, with a few valence particles, are always interesting as they furnish data useful in constructing empirical shell model Hamiltonian consisting of single-particle energies of the valence orbitals and the residual nucleon-nucleon interaction matrix elements. The high-spin states also provide an important opportunity to examine the behavior of multi-particle excitations involving high angular momentum orbits. With increasing excitation energy and angular momentum, certain regularity develops in their spectra due to various underlying reasons. For example, collective bands with moderate deformation as well as superdeformed bands [1] have been reported in these nuclei. Moreover, due to the availability of particles and holes in high-j orbitals, such as $\pi h^{n}_{11/2}$ and $\nu h^{-n}_{11/2}$, phenomena like magnetic rotation could be expected, and have, in fact, been observed in $^{142,143,144}$Gd [2, 3], $^{141,144}$Eu [4, 5] nuclei. These magnetic rotation bands have been interpreted in the light of semi-classical picture as introduced by Macchiavelli *et al.* [6] and more rigorously, in the framework of the tilted axis cranking (TAC) calculation [7].

The present work is part of the attempt for the systematic study of the high spin level structure of nuclei belonging to $A$ ~ 146 region and lying in the vicinity of $N$ = 82 shell closure. In this context, we have investigated the level structure of $^{146}$Gd nucleus. The level structure of $^{146}$Gd has been previously established by several experiments [8-14] up to an excitation energy of $E_x$ ~ 16 MeV and J = 34ħ. The earlier investigations on the multipolarity assignments in $^{146}$Gd were based on the angular correlation and linear polarization measurements [12]. The motivation of the present investigation was to study the high-spin states as well as unambiguous multipolarity assignment of the observed transitions with an array of higher efficiency. We have used a special feature of Clover detector for multipolarity assignments.

## II. EXPERIMETAL PROCEDURE AND DATA ANALYSIS

The excited states in $^{146}$Gd were populated by the $^{115}$In ($^{34}$S, 1$p$2$n$) reaction at a beam energy of 140 MeV. An isotopically-enriched $^{115}$In target of ~ 1.29 mg/cm$^2$ thickness with a ~ 7.14 mg/cm$^2$ Au backing was used. The $^{34}$S beam was provided by the 15UD Pelletron facility at the Inter University Accelerator Centre (IUAC), New Delhi. The γ-rays were detected with INGA [*the Indian National Gamma Array*][15], comprised of 8 Compton-suppressed Clover detectors. The efficiency (after add-back) of the Clover detectors used in the array is typically around 0.17 % (photo-peak) at ~ 1 MeV. These detectors were placed at 80$^{o}$ and 140$^{o}$ with respect to the beam direction. A total of about 220 million two- or higher-fold $\gamma$- $\gamma$ coincidence events were recorded in the experiment.

A systematic procedure has been adopted for detailed off-line analysis. During the pre-sorting of the acquired data, instrumental drifts [16], if any, were corrected for. The add-back spectra were obtained after appropriate gain-matching [17]. From this data, the conventional $E_\gamma$ - $E_\gamma$ histograms (symmetric as well as angle-dependent) were generated. The pre-sorting and sorting were performed using the IUCSORT [18] software. Subsequent data analysis was carried out using both IUCSORT [18] and RADWARE [19] gamma analysis packages.

Multiplicity of the de-exciting $\gamma$ rays from $^{146}$Gd were deduced from the observed $\gamma$ ray angular correlations [20, 21]. For the DCO ratios, $R_{DCO}$, the coincidence data were sorted into an asymmetric matrix whose one axis corresponded to $\gamma$-ray energy deposited in the detectors at 140$^{o}$, while the other axis corresponded to the $\gamma$-ray energy deposited in the detectors at 80$^{o}$. A gate corresponding to a $\gamma$ ray of known multipolarity was made on one axis (say, the x-axis) and the coincidence spectrum was projected on the other axis (y-axis). Next, the same gate was set on the y-axis and projection was made on x-axis. Assuming stretched transitions, the intensities of the transitions that had the same multipolarity as the gated $\gamma$ -ray were approximately the same in both the spectra. On the other hand, if the $\gamma$ rays were of different multipolarity, the intensities differed by a factor of almost two. Using sum gates of known quadrupole transitions, we define $R_{DCO}$ as:

$$\mathrm{R}_{DCO} = \frac{I_{\gamma_1} \text{ at } 80^{o}\text{, gated with } \gamma_2 \text{at } 140^{o}}{I_{\gamma_1} \text{ at } 140^{o}\text{, gated with } \gamma_2 \text{at } 80^{o}}.$$

The extracted values of $R_{DCO}$ are presented in Fig. 1, with detailed information provided in Table I. The $R_{DCO}$ value for known dipole transitions is ~ 0.56 and is ~ 1.05 for known quadruples, when gates are set on quadrupole transitions; using a dipole gating transition, the corresponding values are ~ 1.1 and ~ 1.8 respectively. Further, the $R_{DCO}$ data correctly reproduces the known multipolarities [$\Delta J = 1$ and $\Delta J = 2$] in the neighboring nuclei.

Linear polarization measurement, *i.e.* the direction of electric vector with respect to the beam detector plane, distinguishes the electric and magnetic nature of the associated γ-ray transition. In conjunction with the angular correlation data, these measurements help us to unambiguously assign spins and parities to the observed levels without taking recourse to detailed theoretical calculations. The linear polarization measurements were performed using Clover detector. The ``Clover'' detector has four HPGe crystals arranged in a Clover-leaf configuration. Each individual crystal of the Clover acts as scatterer and the two adjacent crystals as absorbers. With this configuration, the Clover detector behaves as a conventional four-polarimeter in a single cryostat. A useful event for the measurement of linear polarization takes place if the incident γ-ray is Compton scattered from one segment of the clover (scatterer) and the scattered photon is fully absorbed in another segment (analyzer). In this case, the sum of the coincident signals of the two segments gives the energy of the γ -ray while the azimuthal scattering angle from the scatterer to the analyzer detector contains the information on the linear polarization. Multi-Clover array facilitated coincidence polarization measurements [22- 24]. This method has the unique advantage of reducing the contamination from neighboring nuclei and background, thereby resulting in relatively unambiguous assignments for the parity of the states whose spin value is already known. The linear polarization of the radiation can typically be determined (qualitatively) from the difference between the number of Compton scattered γ rays in the reaction plane $N_{parallel}$, and in a plane perpendicular to it, $N_{perp}$.

Figure 2 illustrates the sensitivity of the Clover detector to the linear polarization of the γ rays by comparing the relative intensities of the transitions in the difference spectrum of perpendicular and parallel coincidence following the procedure suggested by Duchene *et al.* [22]. As seen from figure, the 697-, 865-, 886-, and 914-keV transitions are more intense in perpendicular spectrum than in parallel spectrum, which is characteristic of electric transitions. On the other hand, the transitions of 736-, 766-, and 1631-keV are magnetic.

A polarization matrix, using the procedure described in detail in Refs. [23, 24], was constructed from the data corresponding to the energy recorded in any detector on one axis, while the other axis corresponded to the energy scattered in a perpendicular or parallel segment of the Clover with respect to the beam axis. From the projected spectra, the number of perpendicular ($N_{perp}$) and parallel ($N_{parallel}$) scatters for a given γ -ray could be obtained. From these spectra, the asymmetry parameter, $\Delta_{IPDCO}$, was obtained using the relation:

$$\Delta_{IPDCO} = \frac{(a(E_\gamma)N_{perp}) - N_{parallel}}{(a(E_\gamma)N_{perp}) + N_{parallel}}.$$

The correction parameter, *a*, due to the asymmetry of the present experimental configuration has been deduced from radioactive sources ($^{152}$Eu) (having no spin alignment), and is a function of the γ-ray energy, *i.e.* ,

$$a(E_\gamma) = a_0 + a_1 E_{\gamma,}$$

where, $a_0$= 1.016(6) and $a_1$ = -1.32(53) x $10^{-5}$ $(\text{keV})^{-1}$ [25]. In a similar experiment [26], the correction parameter was found constant up to ~ 3 MeV.

A pure electric transition, due to its preferential scattering in the perpendicular direction with respect to the beam axis, results in a positive value for $\Delta_{IPDCO}$. Likewise, a pure magnetic transition results in a negative value for $\Delta_{IPDCO}$ due to its preferential scattering along the parallel direction and a near-zero value for $\Delta_{IPDCO}$ is indicative of a mixed transition. Figure 3 illustrates the results of this procedure and the detailed $\Delta_{IPDCO}$ information is provided in Table I. The present statistics did not permit us to obtain the asymmetry parameters for the weak transitions.

Figure 4 shows a two-dimensional plot of asymmetry parameter $\Delta_{IPDCO}$ vs. angular correlation $R_{DCO}$, as defined above. As seen from the plot, the polarization and multipolarity measurements together can give us a reasonable assignment of the spin and parity for the levels without recourse to detailed theoretical calculations.

## III. EXPERIMETAL RESULTS

The level structure of $^{146}$Gd has been investigated previously in various in-beam γ-ray measurements. The low-spin level structure was investigated primarily with α induced reactions [8, 9, 27]. In the subsequent years, high-spin states in $^{146}$Gd have been studied by several groups using heavy-ion induced reactions [10 - 14]. Weil *et al.* [12] have performed a detailed spin-parity measurement of the high-spin states (up to 30 ħ) in $^{146}$Gd using $^{126}$Te ($^{24}$Mg, 4*n*) $^{146}$Gd reaction and employing a γ-γ coincidence technique. However, they were unable to make unambiguous spin-parity assignments to many of the observed levels. In our present work, thanks to the combination of polarization and γ-ray angular correlation measurements, it has been possible to provide unambiguous assignments of the spins and parities for almost all the observed levels.

The level scheme of $^{146}$Gd obtained from the present investigation is shown in Fig. 5; it is in very good agreement with the earlier works [10, 12]. The quality of the data obtained in our work is shown in the background-subtracted gated spectra presented in Figs.6, 7, and 8.

The polarization data has led to several changes in the previously-reported spin-parity assignments. Weil *et al.* [12] had assigned an M2/E3 multipolarity, with a mixing ratio of $\delta = -0.14(5)$, to the 886-keV transition decaying from a level at $E_x$ = 8917 keV. They estimated a ~ 10% intensity contamination from a transition of identical energy in $^{145}$Eu. In the present experiment, $^{145}$Eu was very weakly populated as compared to $^{146}$Gd. A judicious choice of the gating transitions helped us eliminate any contamination from $^{145}$Eu in the 886 keV transition. The present coincidence angular correlation data indicate that this transition involves a change in angular momentum of $\Delta J = 2$, (Fig. 1) which is in agreement with the measurements of Weil *et al.* [12]. However, the polarization measurements indicate an electric nature for this transition. Figures 2 and 9 show clearly the basis of the present assignment. The gated perpendicular and parallel scattered spectra for a few representative transitions in $^{146}$Gd are depicted in Fig. 9, where one spectrum (parallel scattered) has been shifted by 10 channels for easy visualization. As seen from the figure, the 886-keV transition has a preferential scattering along the perpendicular direction, which signifies its electric nature, whereas, for example, the 312-keV transition is magnetic in nature due to its preferential scattering along the parallel direction.

Further, the present polarization measurements contradict the earlier assignments [12] of the electromagnetic nature for the 766- and 1631-keV transitions de-exciting from the levels at $E_x$ = 7166 keV and $E_x$ = 8031 keV, respectively. The earlier assignment indicated an electric nature for both these transitions, with $\Delta J = 1$ and 2, respectively. The present coincidence angular correlation data confirm these spin assignments. However, as seen from Figs. 2 and 3, both these transitions are magnetic in nature and the 766- and 1631-keV transitions have now been assigned as M1 and M2, respectively. The presence of the 865-keV E1 transition, de-exciting from the level at $E_x$ = 8031 keV (see Figs.1 and 3), corroborates the present multipolarity assignments for the 1631-keV cross-over transition [1631(M2) = 766(M1) + 865(E1)]. Hence, the levels at $E_x$ = 7166, 7567, 8031, and 8917 keV are firmly assigned $J^\pi = 17^+$, $17^+$, $18^-$, and $20^-$, respectively, different from the earlier assignments of $J^\pi = 17^-$, $17^-$, $18^+$, and $20^-$ [12].

The M1 nature of the 291-keV transition from the $E_x$ = 8031-keV level helped us to firmly assign the level at $E_x$ = 7740 keV a $J^\pi = 17^-$ ; this was earlier assigned as $J^\pi = 17^+$ [12]. The spin-parity assignments of the higher-lying levels (with $E_x$ > 7740 keV), which ultimately decay out to the aforesaid level, are dependent on the spin-parity of this level. Hence, although the spin assignments for all the levels remain unchanged, the parity of the levels is reversed in sign from the previously-assigned values in Ref. [12]. The presence of many parallel decay pathways further supports the present assignments. The parity for the level at $E_x$ = 10007 keV extracted from the present measurement corroborates the earlier assignment [12].

A fragmentation of the intensity into several parallel cascades has been observed above the $J^\pi = 29^+$ level at $E_x \sim 13$ MeV. The members of these parallel cascades are clearly seen in the 295-keV gated spectrum (see Fig. 7). These levels have been assigned $J^\pi$ values for the first time. The presence of several high-energy transitions *i.e.* 1704, 1306, 1123 keV) indicate the possible excitation of neutrons above the $N = 82$ core.

Another group of levels, decaying via an 803-keV transition to the 6400-keV level, with $J^\pi = 16^+$, has also been observed in the present study. The background- subtracted coincidence spectrum for these transitions is shown in Fig. 8. There were no previously-reported spin-parity assignments for these transitions. We have been able to assign the multipolarity to all the transitions belonging to this group (see Fig. 8 and Table I).

## IV. SHELL MODEL CALCULATIONS AND DISCUSSION

Shell model calculation with truncated $\pi(gdsh)^{14}$ configurations have been performed for $^{146}$Gd by Eβer *et al.* [28] for states up to $J$ = 16. They used a surface delta plus a modified quadrupole-quadrupole and octupole-octupole interaction. We have also performed shell model calculation for $^{146}$Gd in the $\pi(gdsh)$ valence space using the shell model code OXBASH [29]. Untruncated shell model calculation with 14 active protons in this large single particle basis space was beyond our present computational capacity. So, we have used important particle partitions belonging to $\pi(1g_{7/2}^{6-8}1d_{5/2}^{0-6}1d_{3/2}^{0-4}3s_{1/2}^{0-2}1h_{11/2}^{0-8})$ for the positive-parity states and $\pi(1g_{7/2}^{6-8}1d_{5/2}^{0-6}1d_{3/2}^{0-4}3s_{1/2}^{0-2}1h_{11/2}^{1-5})$ for the negative-parity states. Ground state binding energy (-101.284 MeV) obtained in this calculation compares well with the measured one (-101.585 ± 0.014 MeV) [30]. The maximum angular momentum that can be generated in this model space is $J$ = 32 for both positive- and negative-parity states; the observed maximum spins are $31^+$ and $34^-$. For $J^\pi = 34^-$, neutron excitation by $N$ = *82* core-breaking is essential. In fact, ħω ~ 7 MeV and indeed, as one can expect from the observed level scheme in Fig. 5, this occurs beyond $J^\pi = 16^+$ at $E_x$ = 6.4 MeV. In Figs.10 and 11, we present our shell model results for the positive- and negative-parity states, respectively. Excitation energies of the yrast and some non-yrast positive-parity states up to $J^\pi = 19^+$ are reproduced reasonably in the present calculation with *CW*5082 empirical interaction [32] over the $\pi(gdsh) \otimes \nu(fphi)$ model space. Deviations of the predicted level energies from the experimental values are within 300 keV, except for the $11^+$ level and the ordering of the level sequence is well reproduced. Positive-parity states, including the ground state, are highly configuration mixed.

Excitation energies of the yrast and near-yrast negative parity levels up to $9^-$ are well reproduced with active valence protons only. For all other negative parity states, deviations are larger for higher spins, indicating a need for neutron excitations from the core $^{132}$Sn at above 6-7 MeV excitation. Neutron excitations are also important beyond the yrast $J^\pi = 19^+$ state. In addition, the possibility of proton excitation from the $^{132}$Sn core can not be ruled out altogether at higher excitations. It is to be noted that, whereas the yrast $11^+$ level energy is underestimated by ~ 600 keV in the shell model calculation, the yrast $11^-$ level energy is overestimated by the similar amount. It will be quite interesting to perform large-basis shell model calculations with the possibility of both neutron and proton excitations from the $^{132}$Sn core.

## V. CONCLUSION

The level structure of the even-even nucleus $^{146}$Gd has been studied up to an excitation energy of $E_x$ ~ 15 MeV and J = 34ħ. The spin-parity assignments for most of the observed levels have been made unambiguously by using an array of Clover detectors as a Compton polarimeter. The polarization measurements necessitated some changes in the parities of a few high-lying levels. Also the assignments of multipolarities of a few previously-known levels have been made for the first time. Shell model calculations with restricted proton configurations and no neutron excitations could reproduce reasonably well the observed low-lying level structure. Calculations with the full $\pi(gdsh)^{14}$ configurations may be necessary for better reproduction of the observed low- and moderately-higher-spin states. Further, configurations involving the excitations of proton and neutrons across the $Z = 50$ and $N = 82$ shell-closures would be essential for better reproduction of the observed higher-lying states.

## VI. ACKNOWLEDGMENTS

The authors would like to thank all the participants in the joint National effort to set up the Clover array at IUAC, New Delhi. The help received from the accelerator staff at IUAC as well as from our colleagues at UGC-DAE-CSR, Kolkata and at IUAC, New Delhi is also gratefully acknowledged. This work has been supported in part by the INDO-US, DST-NFS grant (DST-NSF/RPO-017/98) and by the U.S. National Science Foundation (grant Nos. INT-01115336, PHY-4-57210 and PHY07-58100).

**TABLE I. Gamma transition energy ($E_\gamma$) in keV, excitation energy ($E_x$) in keV, initial and final spins for the transition, relative intensity ($I_\gamma$), DCO and IPDCO ratio for the γ-ray transitions and assigned multipolarity in $^{146}$Gd.**

| Eγ (keV) | $E_x$ (keV) | $J_i^\pi \rightarrow J_f^\pi$ | $I_\gamma^a$ | $R_{DCO}$ Gate: Dipole | $R_{DCO}$ Gate: Quadrupole | $\Delta_{IPDCO}$ | Multipolarity |
|---|---|---|---|---|---|---|---|
| 91.5 | 5792.3 | $13^+ \rightarrow 12^+$ | 1.6(0.1) | | dipole | | (M1) |
| 102.4 | 5894.7 | $14^+ \rightarrow 13^+$ | 16.8(0.6) | | 0.53(0.04) | magnetic | M1 |
| 107.8 | 11638.6 | $26^+ \rightarrow 25^-$ | 1.3(0.1) | dipole | | | (E1) |
| 111.2 | 3294.0 | $8^- \rightarrow 8^-$ | 13.1(0.5) | 2.40(0.20) | | | (E2) |
| 124.2 | 6120.7 | $15^+ \rightarrow 14^+$ | 3.2(0.1) | | 0.47(0.05) | magnetic | M1 |
| 130.8 | 7165.7 | $17^+ \rightarrow 16^-$ | 6.1(0.2) | 0.86(0.07) | | | (E1) |
| 134.7 | 3428.9 | $9^- \rightarrow 8^-$ | 48.8(1.6) | 1.02(0.06) | | magnetic | M1 |
| 163 | 4178.2 | $(32^-) \rightarrow 31^-$ | 0.7(0.1) | | dipole | magnetic | (M1) |
| 173.5 | 9257.9 | $21^+ \rightarrow 20^-$ | 5.0(0.2) | | 0.56(0.04) | 0.10(0.13) | E1 |
| 197.8 | 1639.1 | $26^+ \rightarrow 25^+$ | 3.3(0.7) | 1.39(0.17) | | magnetic | M1 |
| 200.4 | 3182.8 | $8^- \rightarrow 7^-$ | 31.6(1.2) | 1.04(0.07) | | magnetic | M1 |
| 226 | 6120.7 | $15^+ \rightarrow 14^+$ | 7.2(1.8)[b] | | 0.56(0.03) | -.21(0.13) | M1 |
| 226.3 | 7740.1 | $17^- \rightarrow 16^+$ | | | 0.35(0.03) | electric | E1 |
| 256.4 | 9482.9 | $22^-\} \rightarrow 21^-$ | 3.9(0.2) | 1.16(0.12) | | magnetic | M1 |
| 260.3 | 8000.4 | $18^- \rightarrow 17^-$ | 0.8(0.1) | | 0.40(0.03) | -0.02(0.03) | M1 |
| 270.2 | 9528.1 | $22^+ \rightarrow 21^+$ | 3.8(0.2) | | 0.55(0.03) | -0.09(0.10) | M1 |
| 278.9 | 6399.6 | $16^+ \rightarrow 15^+$ | 30.3(1.0) | | 0.56(0.03) | -0.03(0.02) | M1 |
| 291.3 | 8031.4 | $18^- \rightarrow 17^-$ | 5.5(0.2) | 0.91(0.06) | | -0.10(0.06) | M1 |
| 295.2 | 11933.8 | $27^+ \rightarrow 26^+$ | 9.1(0.4) | | 0.49(0.03) | -0.09(0.10) | M1 |
| 297.5 | 8666.9 | $19^- \rightarrow 18^-$ | 7.2(0.3) | | 0.56(0.04) | -0.06(0.06) | M1 |
| 310 | 9226.5 | $21^- \rightarrow 20^-$ | | 1.01(0.07) | | -0.04(0.03) | M1 |
| 311.8 | 3294.2 | $8^- \rightarrow 7^-$ | 68.7(2.2)[c] | 1.13(0.08) | | -0.06(0.05) | M1 |
| 315.9 | 15287.3 | $33^- \rightarrow 32^-$ | 0.8(0.1) | | 0.67(0.06) | magnetic | M1 |
| 324.1 | 2982.4 | $7^- \rightarrow 5^-$ | 94.8(3.1) | 1.97(0.13) | | 0.06(0.04) | E2 |
| 328.4 | 6120.7 | $15^+ \rightarrow 13^+$ | 3.2(0.2) | | 1.03(0.08) | 0.27(0.07) | E2 |
| 343.8 | 5792.3 | $13^+ \rightarrow 12^+$ | 1.0(0.1) | 1.09(0.09) | | -0.11(0.10) | M1 |
| 375.2 | 14971.4 | $32^- \rightarrow 31^+$ | 0.8(0.1) | | 0.49(0.07) | 0.19(0.11) | E1 |
| 393.6 | 11639.1 | $26^+ \rightarrow 25^-$ | 2.3(0.1) | 0.95(0.07) | | 0.07(0.08) | E1 |
| 403.1 | 6399.6 | $16^+ \rightarrow 14^+$ | 0.7(0.1) | | 1.01(0.09) | | (E2) |
| 416.1 | 11441.3 | $25^+ \rightarrow 24^+$ | | dipole | | magnetic | (M1) |
| 417.5 | 9084.4 | $20^- \rightarrow 19^-$ | 6.4(0.3)[d] | 0.78(0.04) | | magnetic | M1 |
| 434 | 9084.4 | $20^- \rightarrow 19^+$ | 1.1(0.1) | 0.77(0.09) | | electric | E1 |
| 436.4 | 3865.3 | $10^+ \rightarrow 9^-$ | 73.2(2.4) | 1.03(0.06) | | 0.10(0.07) | E1 |
| 441.2 | 5792.4 | $13^+ \rightarrow 12^+$ | 3.0(0.1) | | 0.47(0.03) | -0.03(0.03) | M1 |
| 446.1 | 5895.0 | $14^+ \rightarrow 12^+$ | 4.6(0.2) | | 0.97(0.07) | 0.24(0.06) | E2 |
| 463.9 | 8030.7 | $18^- \rightarrow 17^+$ | 1.1(0.1) | | 0.57(0.04) | 0.11(0.09) | E1 |
| 479.2 | 10007.3 | $23^- \rightarrow 22^+$ | 2.1(0.1) | | 0.47(0.03) | 0.21(0.11) | E1 |
| 504.9 | 6399.6 | $16^+ \rightarrow 14^+$ | | | quadrupole | electric | E2 |
| 505.6 | 11530.8 | $25^- \rightarrow 24^+$ | 12.0(0.4)[e] | 1.25(0.10) | | 0.05(0.04) | E1 |
| 514.3 | 5792.2 | $13^+ \rightarrow 11^+$ | 11.6(0.4) | 2.21(0.18) | | 0.14(0.05) | E2 |
| 543.6 | 5894.8 | $14^+ \rightarrow 12^+$ | 2.3(0.1) | | 1.24(0.09) | 0.07(0.08) | E2 |
| 554.4 | 15841.7 | $34^- \rightarrow 33^-$ | 0.7(0.1) | | 0.60(0.08) | -0.10(0.11) | M1 |
| 592.8 | 5095.3 | $11^+ \rightarrow 10^+$ | 8.4(0.3) | 0.72(0.05) | | -0.11(0.06) | M1 |
| 629.3 | 8369.4 | $18^- \rightarrow 17^-$ | 0.7(0.1) | 1.09(0.13) | | magnetic | M1 |
| 645.3 | 5996.5 | $14^+ \rightarrow 12^+$ | 6.9(0.3) | | 1.05(0.07) | 0.09(0.08) | E2 |
| 650.0 | 8650.4 | $19^+ \rightarrow 18^-$ | 1.2(0.1) | 0.86(0.05) | | 0.20(0.12) | E1 |
| 658.6 | 11100.2 | $23^- \rightarrow 22^-$ | 0.3(0.1) | dipole | | | |
| 669.7 | 11441.2 | $25^+ \rightarrow 24^-$ | 6.0(0.2) | | 0.51(0.05) | electric | E1 |
| 690.9 | 9496.4 | $20^- \rightarrow 19^-$ | 0.4(0.1) | | | | |
| 695.0 | 10441.6 | $22^- \to 21^-$ | 0.8(0.1) | | 0.41(0.11) | magnetic | M1 |
| 697.0 | 5792.3 | $13^+ \to 11^+$ | 28.1(1.0) | | 1.08(0.06) | 0.08(0.04) | E2 |
| 709.1 | 8369.0 | $18^- \to (17^+)$ | 1.9(0.1) | dipole | | electric | (E1) |
| 727.4 | 8805.5 | $19^- \to 18^-$ | 3.7(0.2) | 0.80(0.12) | | -0.10(0.08) | M1 |
| 735.9 | 5278.1 | $11^+ \to 10^+$ | 5.0(0.2) | | 0.58(0.05) | magnetic | M1 |
| 757.4 | 11025.1 | $24^+ \to 23^-$ | 1.4(0.2) | | 0.46(0.07) | electric | E1 |
| 766.0 | 7165.6 | $17^+ \to 16^+$ | 18.9(0.7) | | 0.50(0.03) | -0.06(0.05) | M1 |
| 780.9 | 4646.2 | $11^- \to 10^+$ | 2.6(0.1) | 0.96(0.07) | | 0.16(0.06) | E1 |
| 802.7 | 5448.9 | $12^+ \to 11^-$ | 2.2(0.1)[f] | dipole | | electric | (E1) |
| 803.0 | 7202.6 | $17^- \to 16^+$ | | 1.14(0.15) | | 0.19(0.12) | E1 |
| 805.6 | 13698.2 | $31^+ \to 29^+$ | 3.0(0.2) | | 1.26(0.13) | 0.24(0.08) | E2 |
| 807.7 | 10771.5 | $24^- \to 22^-$ | 4.2(0.2) | | quadrupole | electric | (E2) |
| 861.7 | 10088.2 | $23^- \to 21^-$ | 6.1(0.3) | 1.76(0.15) | | 0.19(0.13) | E2 |
| 865.2 | 8030.8 | $18^- \to 17^+$ | 23.9(0.8) | | 0.57(0.03) | 0.11(0.07) | E1 |
| 875.5 | 8078.1 | $18^- \to 17^-$ | 1.7(0.1) | | 0.65(0.11) | -0.12(0.11) | M1 |
| 885.7 | 8916.5 | $20^- \to 18^-$ | 35.1(1.2) | | 1.09(0.06) | 0.06(0.04) | E2 |
| 914.2 | 7034.9 | $16^- \to 15^+$ | 10.7(0.4) | | 0.59(0.04) | 0.11(0.05) | E1 |
| 937.0 | 11025.2 | $24^+ \to 23^-$ | 5.8(0.2) | | 0.51(0.04) | 0.12(0.12) | E1 |
| 941.1 | 9746.6 | $21^- \to 19^-$ | 1.5(0.1) | 1.70(0.24) | | electric | E2 |
| 945.2 | 10441.6 | $22^- \to 20^-$ | 0.7(0.1) | 1.86(0.23) | | electric | E2 |
| 958.8 | 12892.6 | $29^+ \to 27^+$ | 8.0(0.3) | | 1.06(0.08) | 0.12(0.08) | E2 |
| 977.8 | 11245.5 | $25^- \to 23^-$ | 4.8(0.2) | 1.76(0.17) | | 0.09(0.05) | E2 |
| 1041.2 | 10267.7 | $23^- \to 21^-$ | 10.8(0.4) | | 1.17(0.08) | 0.08(0.05) | E2 |
| 1047.3 | 9963.8 | $22^- \to 20^-$ | 8.5(0.3) | | 1.14(0.08) | 0.07(0.04) | E2 |
| 1073.6 | 4502.5 | $10^+ \to 9^-$ | 8.5(0.3) | 0.97(0.08) | | 0.07(0.08) | E1 |
| 1078.8 | 2658.3 | $5^- \to 3^-$ | 100.0 | | 1.1(0.7) | 0.04(0.02) | E2 |
| 1113.3 | 4542.2 | $10^+ \to 9^-$ | | 1.10(0.08) | | electric | E1 |
| 1114.2 | 7513.8 | $16^+ \to 16^+$ | 17.4(0.6)[g] | | quadrupole | electric | (E2) |
| 1122.6 | 14015.2 | $31^- \to 29^+$ | 1.7(0.1) | | 0.92(0.14) | magnetic | M2 |
| 1167.2 | 7566.8 | $17^+ \to 16^+$ | 5.8(0.2) | 1.03(0.07) | | -0.10(0.08) | M1 |
| 1230.0 | 5095.3 | $11^+ \to 10^+$ | 25.6(0.9) | | 0.48(0.03) | -0.08(0.07) | M1 |
| 1260.3 | 7659.9 | $(17^+) \to 16^+$ | 0.3(0.1) | | | | |
| 1288.6 | 10771.5 | $24^- \to 22^-$ | 6.2(0.3) | | 0.89(0.06) | 0.08(0.05) | E2 |
| 1305.6 | 14198.2 | $(30^+) \to 29^+$ | 2.0(0.1) | dipole | | magnetic | (M1) |
| 1412.6 | 5277.9 | $11^+ \to 10^+$ | 5.5(0.2) | | 0.56(0.04) | -0.17(0.09) | M1 |
| 1417.2 | 2996.7 | $(5^-) \to 3^-$ | 0.5(0.1) | quadrupole | | | (E2) |
| 1485.9 | 5351.2 | $12^+ \to 10^+$ | 24.8(0.8) | | 1.17(0.08) | 0.09(0.06) | E2 |
| 1579.5 | 1579.5 | $3^- \to 0^+$ | >105 | 1.38(0.08) | | 0.07(0.06) | E3 |
| 1583.2 | 5448.5 | $12^+ \to 10^+$ | 5.0(0.3) | 1.82(0.19) | | 0.08(0.9) | E2 |
| 1631.3 | 8030.9 | $18^- \to 16^+$ | 3.3(0.2) | 2.07(0.19) | | -0.12(0.09) | M2 |
| 1703.6 | 14596.2 | $31^+ \to 29^+$ | 0.9(0.1) | | 0.93(0.15) | 0.18(0.15) | E2 |
| 1835.5 | 5700.8 | $12^+ \to 10^+$ | 5.0(0.2) | | 1.21(0.09) | electric | E2 |
| 1865.1 | 5730.4 | $(12^+) \to 10^+$ | 3.1(0.2) | | quadrupole | electric | (E2) |
| 2061.1 | 4719.4 | $(7^-) \to 5^-$ | 1.2(0.1) | | quadrupole | | (E2) |
| 2327.6 | 3907.1 | $(5^-) \to 3^-$ | 1.8(0.1) | quadrupole | | | (E2) |
| 2627.6 | 4207.1 | $(5^-) \to 3^-$ | 2.2(0.1) | quadrupole | | | (E2) |

[a] The quoted errors on intensities encompass errors due to background subtraction, peak fitting and efficiency correction.
[b] Combined intensity of 226.0 and 226.3 keV transitions.

[c] Combined intensity of 310.0 and 311.8 keV transitions.
[d] Combined intensity of 416.1 and 417.5 keV transitions.
[e] Combined intensity of 504.9 and 505.6 keV transitions.
[f] Combined intensity of 802.7 and 803.0 keV transitions.
[g] Combined intensity of 1113.3 and 1114.2 keV transitions.

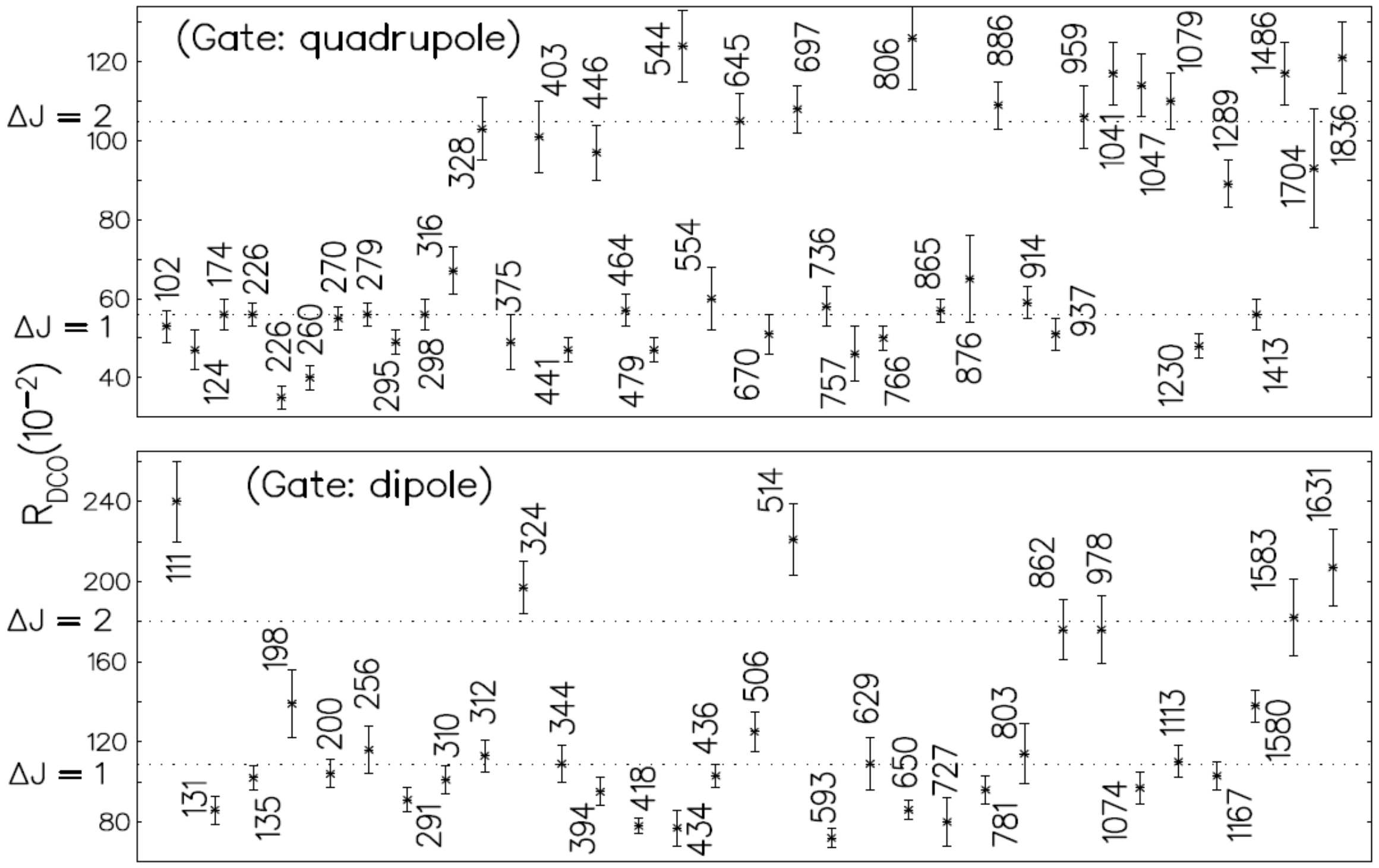

**FIG. 1 γ-ray anisotropy intensity ratio, $R_{DCO}$, for the transitions belonging to $^{146}$Gd. The dotted lines correspond to the average values obtained for known dipole ΔJ = 1 and quadrupole ΔJ = 2 transitions when gated by quadrupole (top panel) and dipole (bottom panel) transitions. The quoted errors include the uncertainties associated with background subtraction, peak fitting, and efficiency correction.**

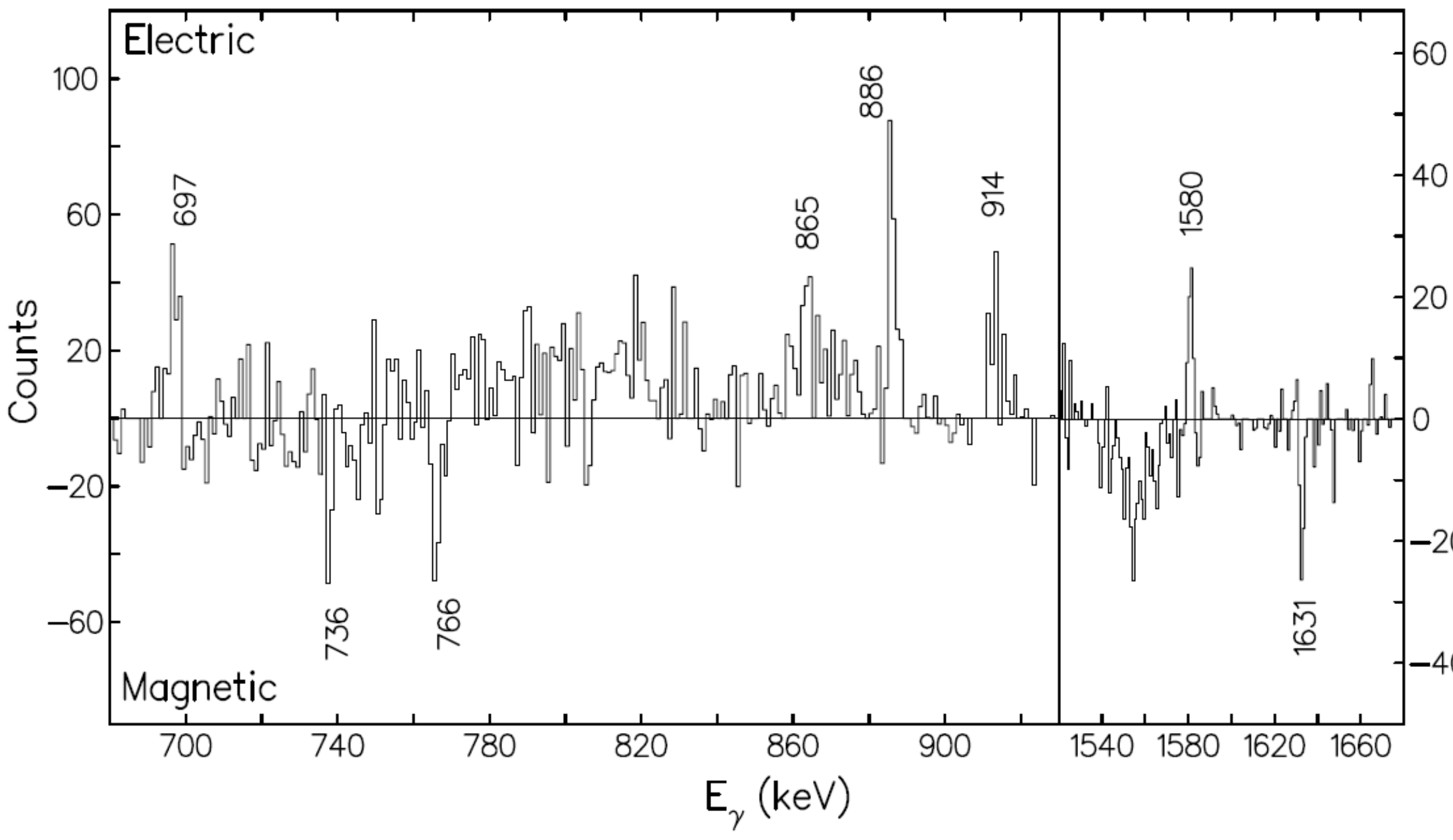

**FIG.2 Background-subtracted difference spectrum (perpendicular minus parallel coincidences). The electric γ-ray transitions result in positive peaks, whereas the magnetic transitions lead to negative peaks.**

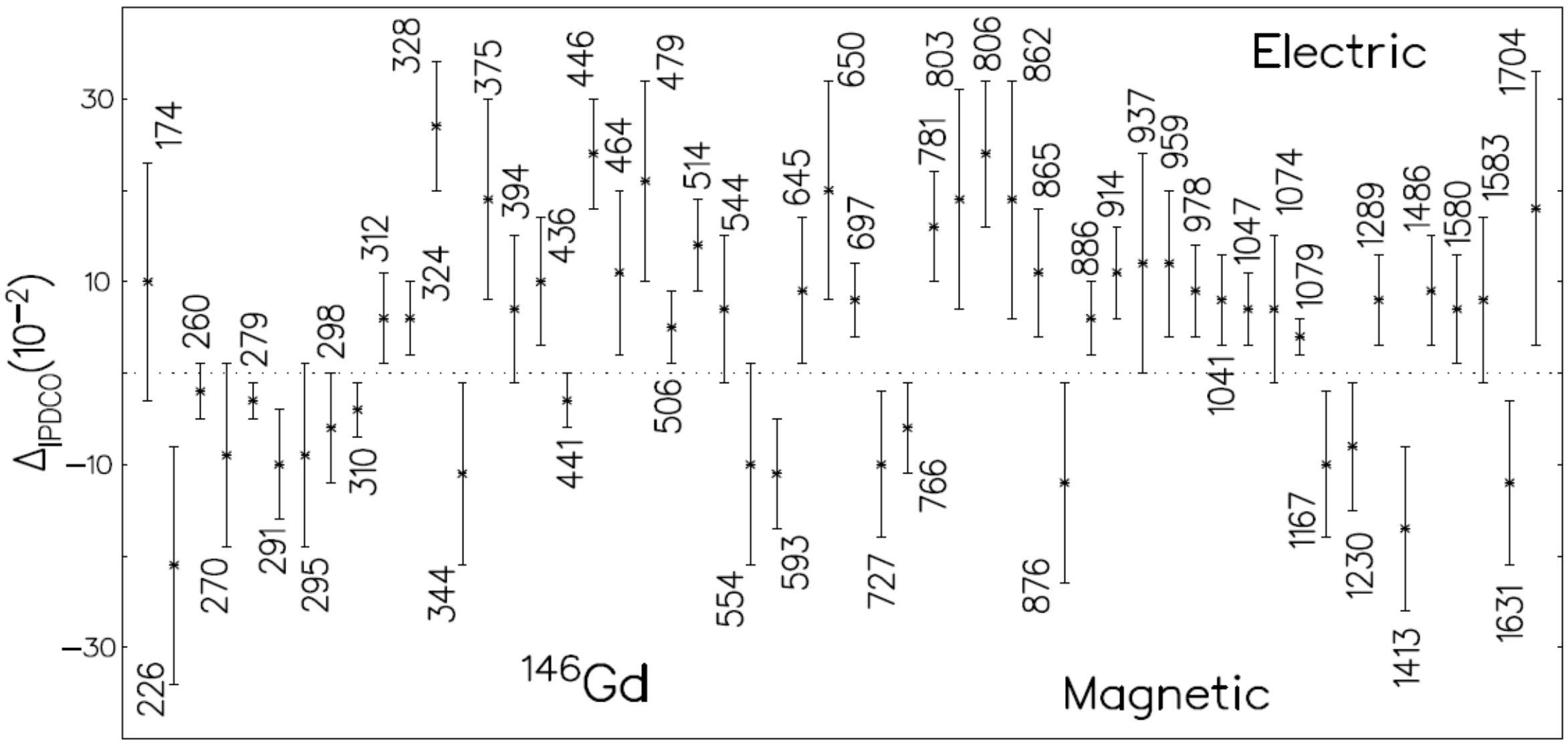

**FIG. 3 Representative experimental γ-ray asymmetry parameters, $\Delta_{IPDCO}$, from polarization measurements for selected γ-ray transitions of $^{146}$Gd. A positive value corresponds to an electric transition, while a magnetic transition results in a negative value. The quoted error encompasses uncertainties due to background subtraction and fitting. The dotted line indicates the zero value of $\Delta_{IPDCO}$ and has been drawn to guide the eye**

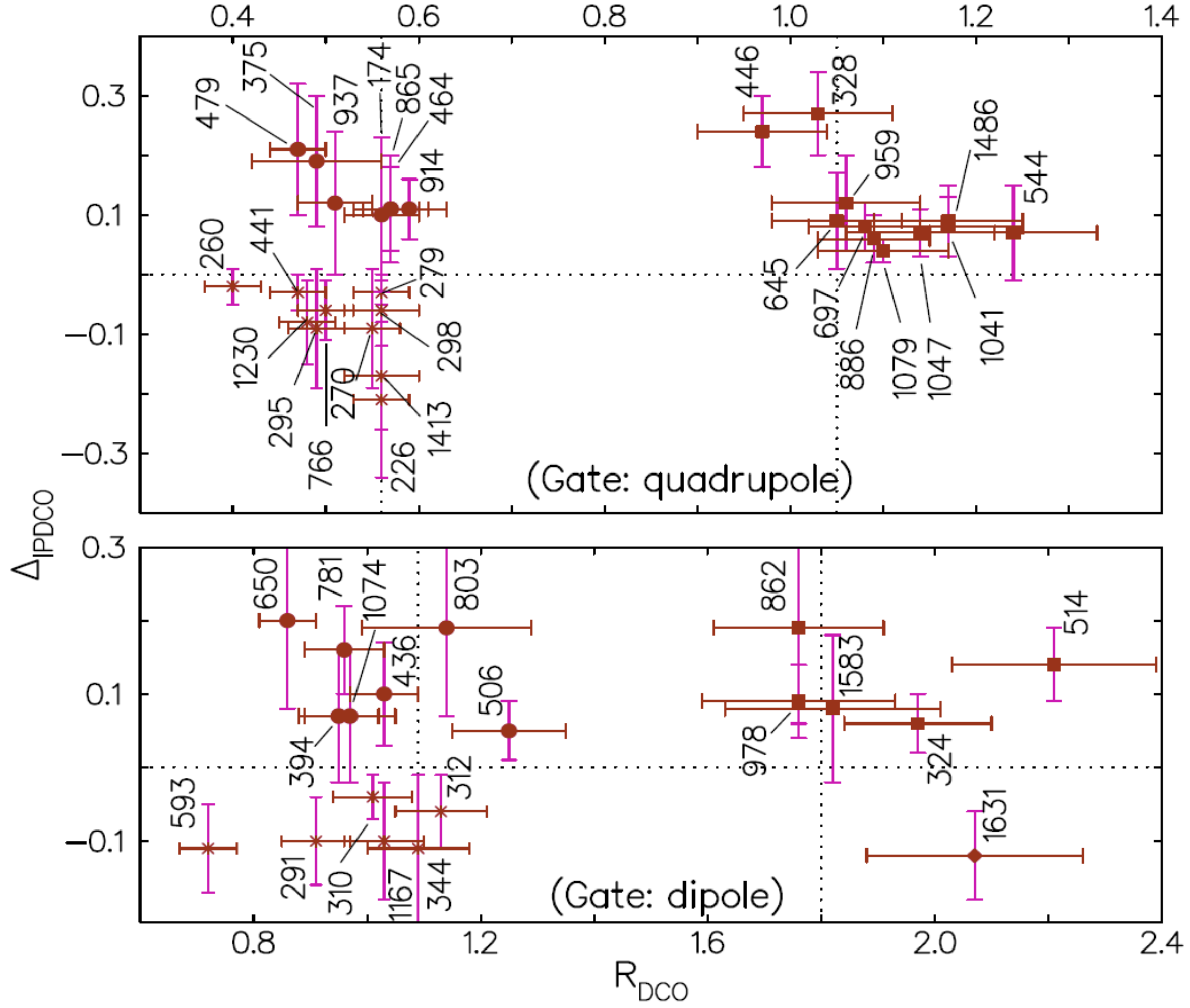

**FIG. 4 ``Color online'' Two-dimensional plot of the asymmetry parameter $\Delta_{IPDCO}$ vs. the angular correlation ratio $R_{DCO}$. Two types of transitions (electric and magnetic) create separate clusters for E1 (filled circles), M1 (asterisks), E2 (filled squares) and M2 (filled diamond) multipolarities. The dotted lines parallel to the y-axis correspond to the value obtained for known dipole and quadrupole transitions, respectively (gated on both dipoles and quadruples). The dotted line parallel to the x-axis indicates the zero value of $\Delta_{IPDCO}$.**

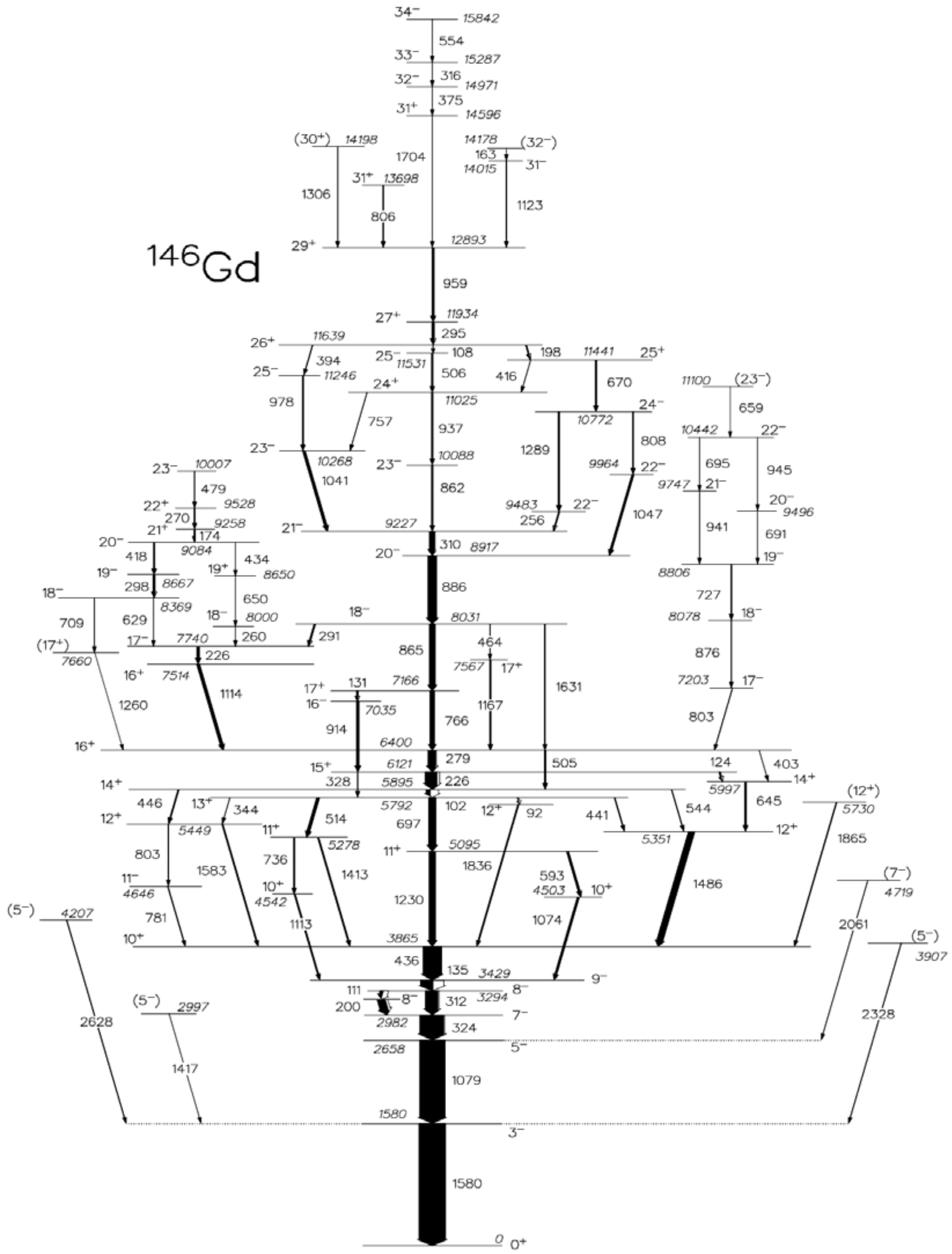

**FIG. 5 The level scheme of $^{146}$Gd as obtained from the present investigation. The width of the arrows approximately represents the observed intensities.**

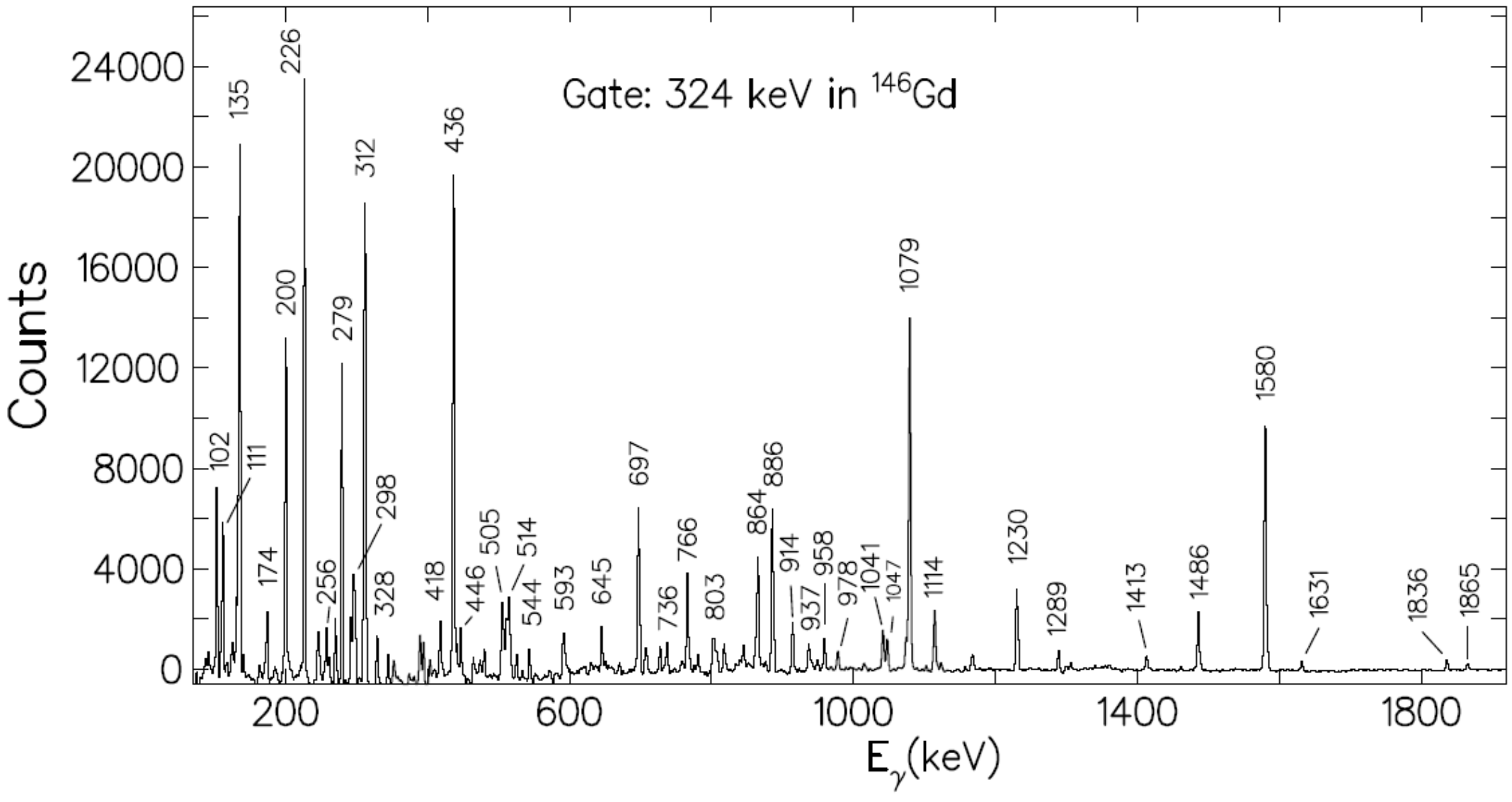

**FIG. 6 Background-subtracted spectrum gated on the 324-keV transition in $^{146}$Gd. Peaks associated with $^{146}$Gd are labeled with their energies in keV**.

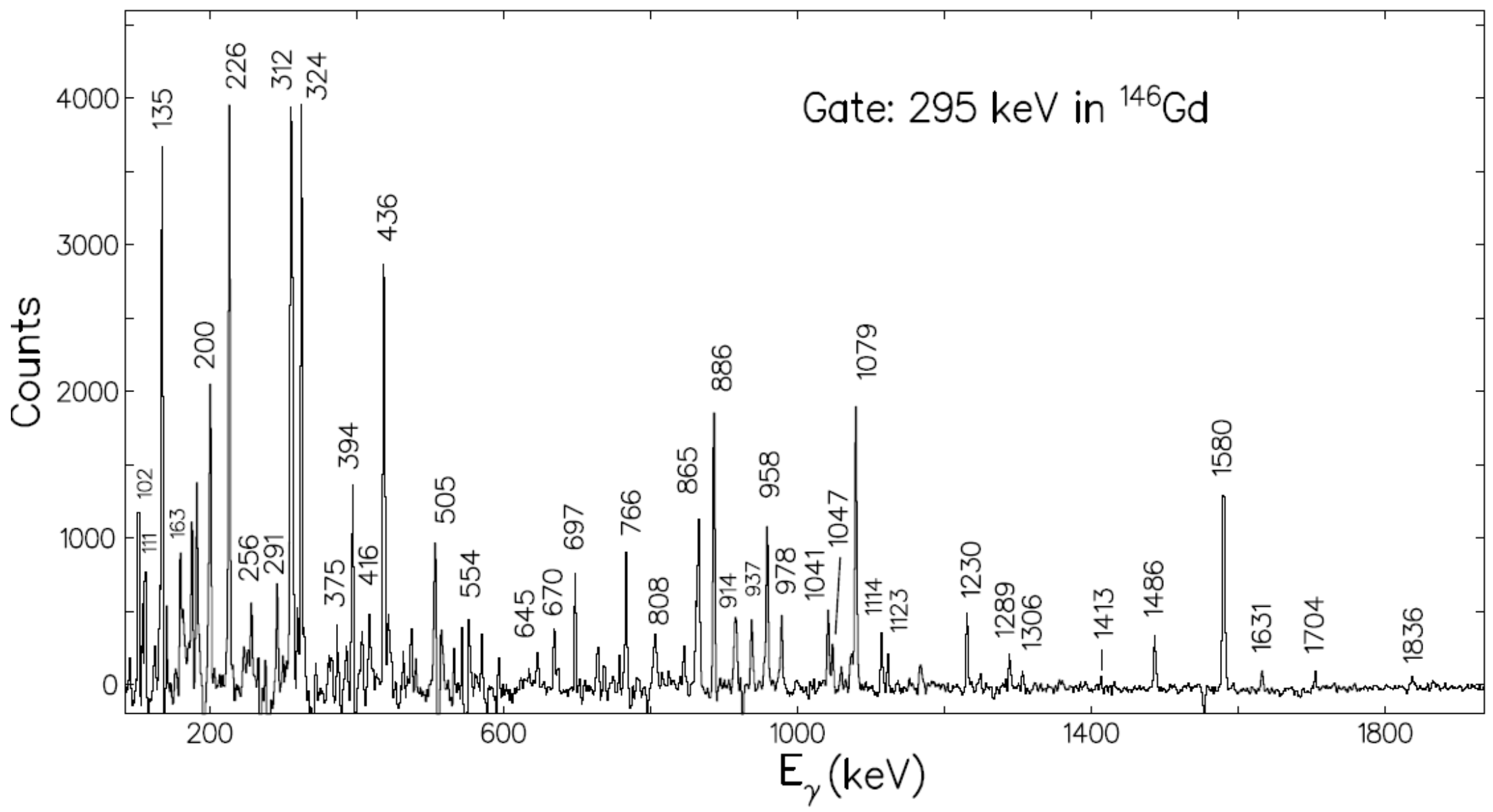

**FIG. 7 Background-subtracted spectrum gated on the 295-keV transition in $^{146}$Gd. Peaks associated with $^{146}$Gd are labeled with their energies in keV.**

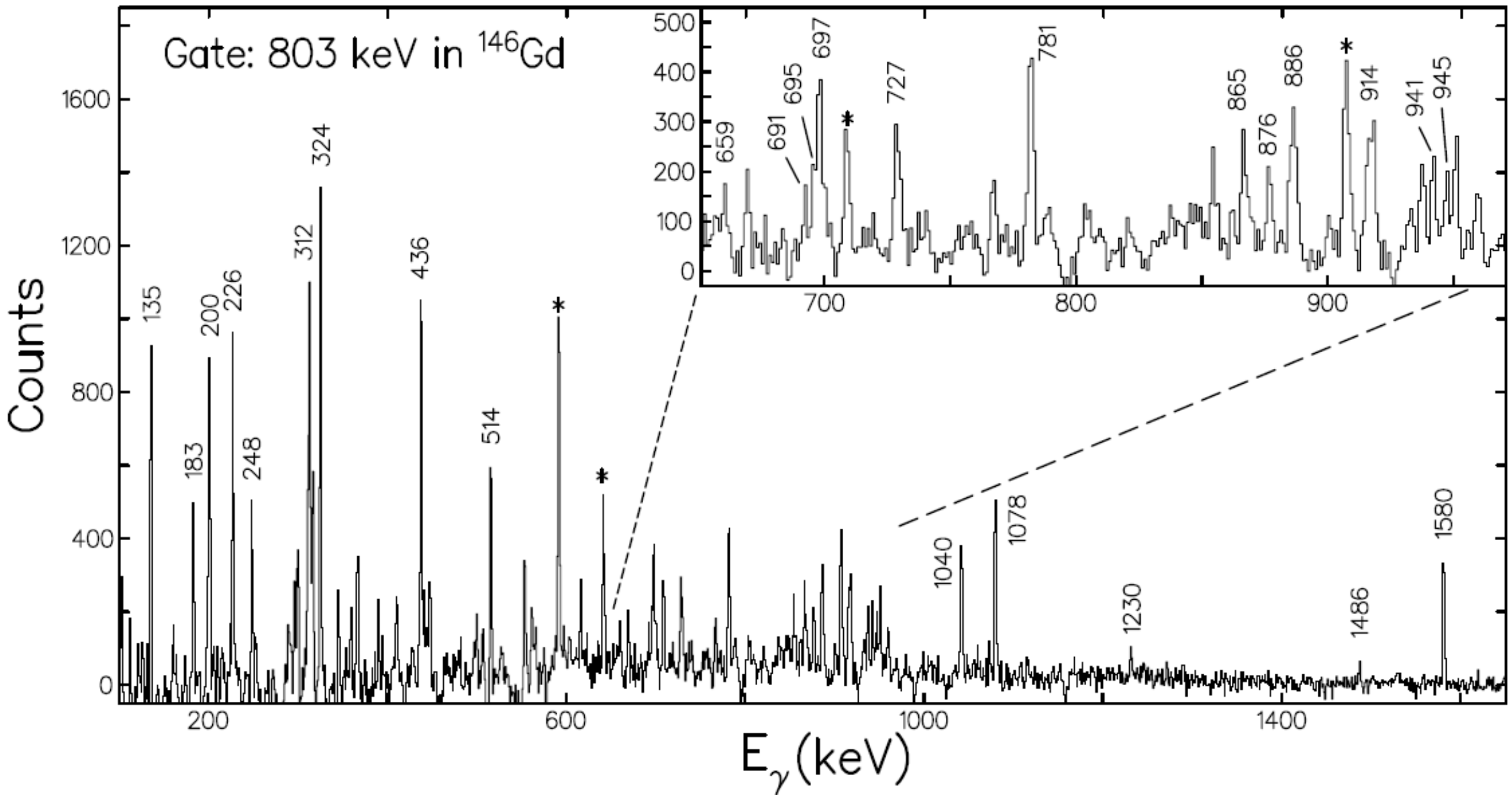

**FIG. 8 Background-subtracted spectrum gated on the 803-keV transition in $^{146}$Gd. Peaks associated with $^{146}$Gd are labeled with their energies in keV. A part of the spectrum has been expanded for clarity. The transitions marked with an asterisk belong to neighboring nuclei.**

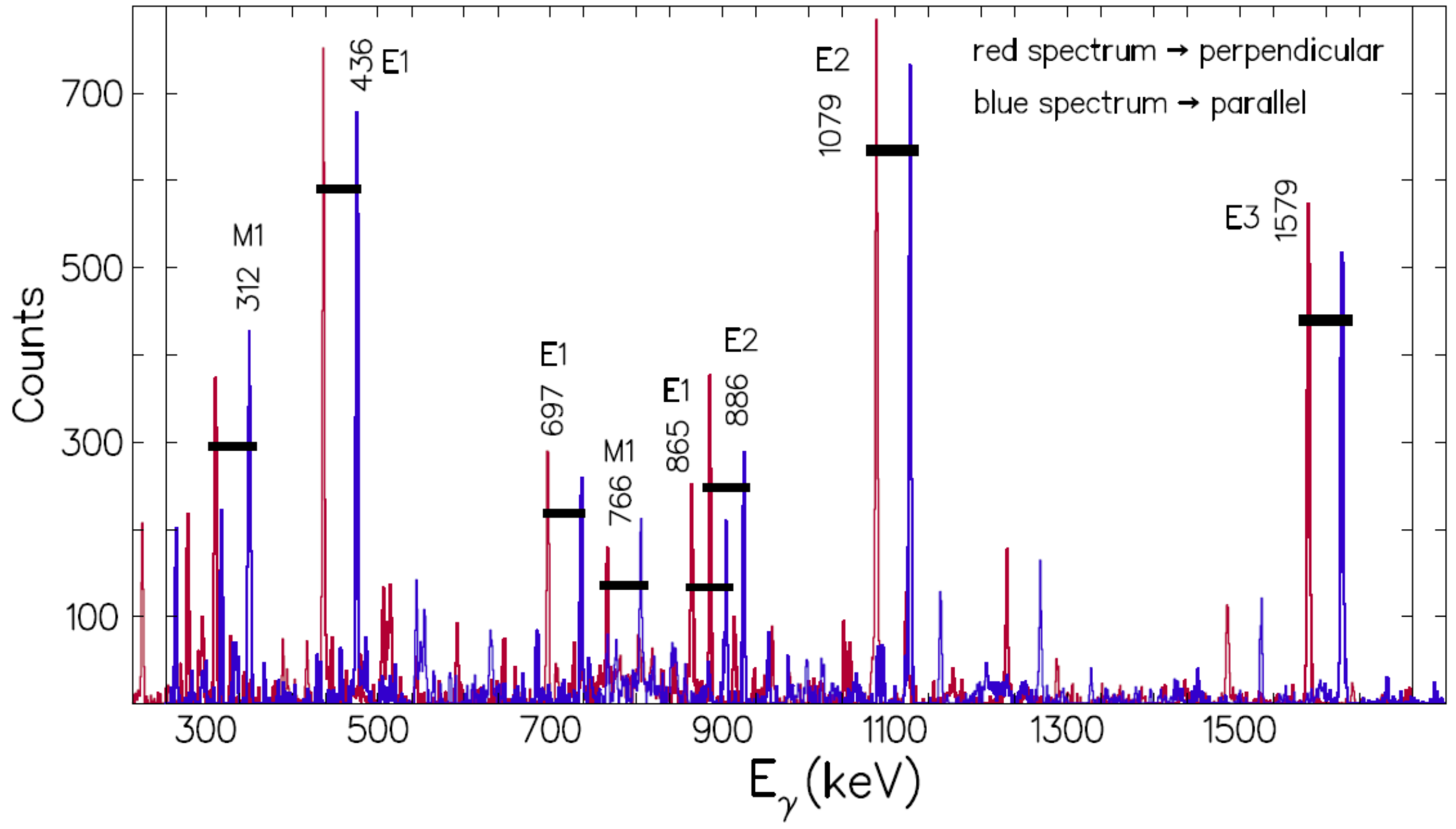

**FIG. 9 ``Color online'' Comparison of parallel and perpendicular scattered spectra with gate on the 324-keV ($7^- \rightarrow 5^-$) transition in $^{146}$Gd. The parallel spectrum (blue colored) has been shifted by 10 channels for better clarity.**

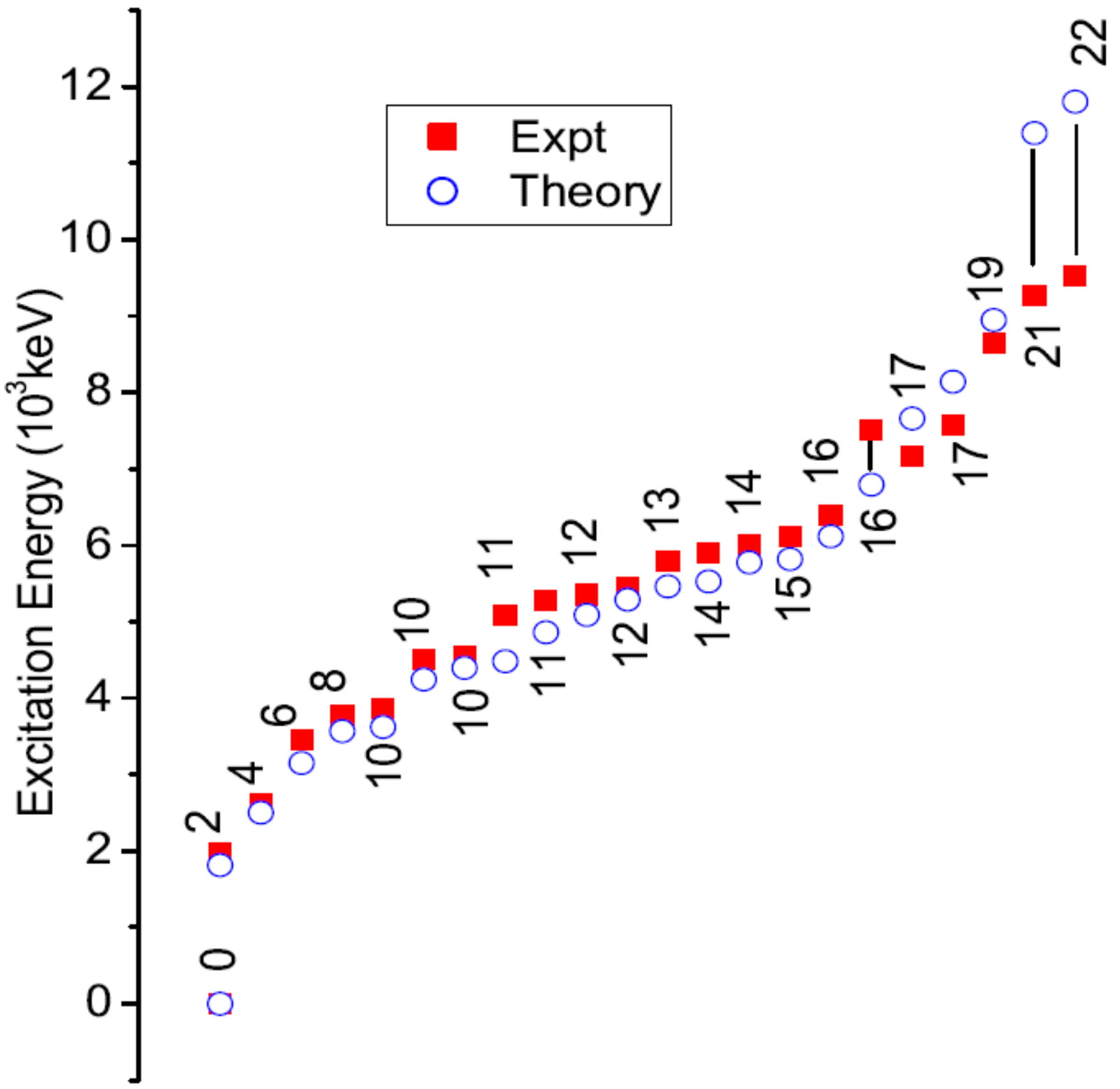

**FIG. 10 ``Color online'' Comparison between the observed and calculated states for the positive parity. The value written on each of the data point in the plot corresponds to the spin of the respective level. The $2^+$, $4^+$, $6^+$, and $8^+$ states have been incorporated from the data base of Ref. [31].**

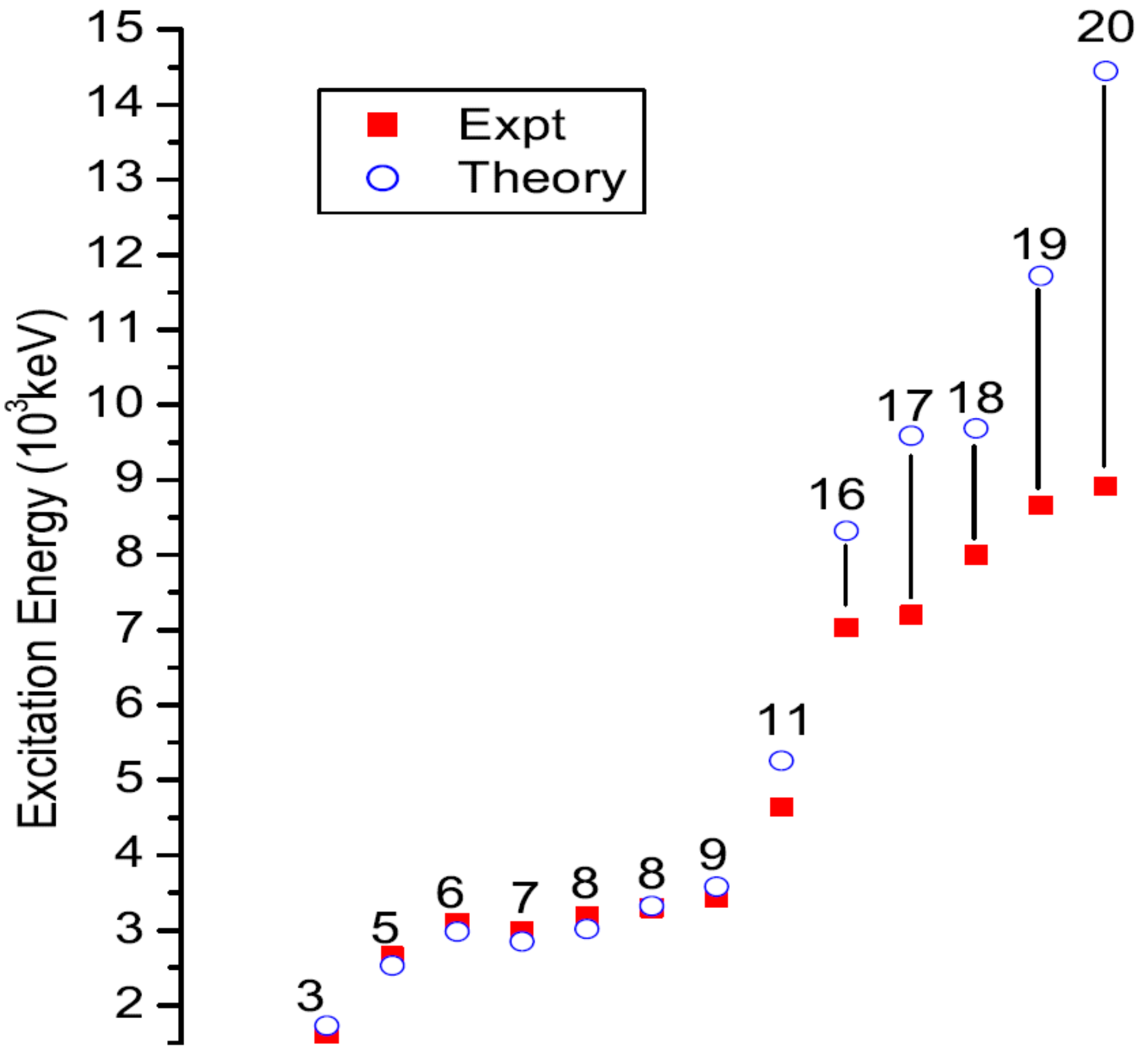

**FIG. 11 ``Color online'' Same as in Fig. 10 but for the negative parity states. The $6^-$ level has been taken from the data base of Ref. [31].**